# CONSUMER RIGHTS AND ALGORITHMS

Gregory M. Dickinson[*]

*This article summarizes the field of consumer protection law, from its historical roots to the contemporary challenges of the digital age. It outlines the legal doctrines governing consumer deception and unfair trade practices, highlighting the interplay between common-law, statutory, and private modes of regulation. The article then addresses the impact of artificial intelligence and big data on consumer markets, focusing on digital advertising and new forms of consumer fraud. Finally, it explores regulatory responses to these challenges, including data privacy laws and prohibitions on dark patterns, which illustrate the trade-offs inherent in consumer protection frameworks.*

## I. THE GUIDEPOSTS OF CONSUMER-PROTECTION LAW

For as long as there have been markets and buyers there have been shysters and charlatans. Consumer law is built on endless scuffles between buyers and sellers, each trying to win for herself the best possible bargain, sometimes acting underhandedly to do so. The earliest reported English cases, dating from the fifteenth century, recount disputes between merchants over goods typical of the time—undelivered sacks of wool, casks of spoiled wine, and other such problems that strike the modern mind as quaint yet relatable. From medieval merchants to con artists of the industrial revolution and door-to-door salesmen of the twentieth century, history shows sellers pitching their products in the best possible light, sometimes even bending or breaking the truth as they do so. (Dickinson [2024] at 2464).

Despite dramatic social and technological change, the law's response across the centuries has remained constant too: Sellers are held to a standard of honesty and fair dealing and must not deceive their customers, while buyers have a complementary obligation to look out for their own interests. That the basic premises of consumer protection law remain constant is notable, but unsurprising, for the problem at issue is the usurpation of the consumer's autonomy by deception or coercion. (Epstein [1980] at 256). The particular strategies employed or the goods or services transacted for are irrelevant.

---

[*] Assistant Professor of Law and, by courtesy, Computer Science, the University of Nebraska College of Law; Nonresidential Fellow, Stanford Law School, Program in Law, Science & Technology; J.D., Harvard Law School. Thanks to Mixa Hernandez for her routinely excellent research assistance.

Even today, the basic principles remain undisturbed. In the United States, for example, the common law of fraud and state Unfair and Deceptive Acts and Practices statutes prohibit sellers from making "material misrepresentations" regarding their products, but only where such misrepresentations are likely to mislead a reasonable consumer. Similarly, in the European Union (EU), the Unfair Commercial Practices Directive prohibits "misleading commercial practices" that are "likely to deceive the average consumer," who is "reasonably well-informed and reasonably observant and circumspect." Across time and geography, marketplace deception is prohibited, but consumers are also expected to look after their own interests.

The traditional rule of caveat emptor, which grounded both Roman and English law, placed the remedial power in the hands of the consumer (Buckland and Arnold [1952] at 284). A buyer was responsible for inspecting the goods on offer; but caveat emptor did not excuse dishonesty. A seller who intentionally misled a buyer or reneged on an agreed-upon deal would be subject to a legal action for fraud or breach of contract. The dual obligations of caveat emptor and honest dealing enable the market to function smoothly, with parties free to negotiate voluntary, mutually beneficial transactions, while reputational considerations encourage fair dealing, and a backstop of legal remedies for fraud prevents the parties from cheating the process through deception or coercion.

Reputation-based market constraints and the threat of private litigation remain key components of the modern consumer-protection system, but they are now bolstered by governmental regulation and enforcement regimes designed to complement them.

One core focus of governmental enforcement efforts is consumer deception,[1] broadly understood to include such practices as false advertising, coercive sales tactics, failure to deliver or delivery of inferior goods, and breaches of product warranties. Sellers' concern to protect their reputations (to encourage future sales) and the background threat of litigation do the lion's share of the work in deterring such underhanded practices. Yet those mechanisms are usefully supplemented by governmental enforcement where a disparity in information between buyer and seller prevents buyers from making informed purchase decisions, (Nelson [1974] at 749), where litigation is too costly to provide efficient redress to consumers, (Fitzpatrick [2019] at 15), and, more controversially, where cognitive biases may prevent consumers from rationally looking after their own interests (Sunstein and Thaler [2021] at 48).

---

[1] The other core focus is antitrust law, on which see generally (Posner [1990] at chap. 1).

## II. THE NEW ERA OF DECEPTION

Over the last three decades, digital technologies have revolutionized commerce; the same tools have also produced a revolution in online fraud. Using enormous collections of historical browsing and shopping data, companies construct detailed consumer profiles, which they then deploy to create products and online shopping experiences responsive to the preferences of their target audience. (Dickinson [2024] at 2472).

The widespread availability of AI algorithms and consumer-information databases has changed the historical relationship between buyer and seller. Traditional product development and marketing are expensive. With automated data collection and AI-powered analysis, however, the product development process can be streamlined and conducted at a fraction of the cost.

Historical software usage data might show, for example, a user-interface bottleneck where a repeated task that requires three mouse clicks could be reduced to a single click by relocating a button to an app's main screen. Or consumer search data might reveal an unmet consumer desire for handbags in "fondant pink" or "terracotta brown." Instead of relying on customer interviews and satisfaction surveys, a business can now discern consumer preferences less expensively and more accurately by analyzing her historical shopping data. (Dickinson [2024]).

AI lowers the cost of advertising too. Traditionally, a new product launch might have been accompanied by an advertising blitz in newspapers or on television. Ads for fancy watches and vacation resorts would appear in the Wall Street Journal and commercials promoting beer or athletic gear during live sporting events. Precise targeting was difficult, however, and mass advertising is expensive. By contrast, AI-driven digital advertising allows precise targeting of ads to those consumers most likely to be interested in a particular product. (Dickinson [2024]).

For sellers, who can purchase only the ads they need, all of this means cheaper advertising campaigns with higher customer realization rates. For consumers, the effects are the same, just from the opposite perspective. Detailed consumer data allows companies to develop products and services that more closely match their preferences and then to tailor advertising campaigns to reach and inform those consumers most likely to be interested. The end results are web-browsing and in-app experiences with more relevant ads, tailored brand experiences, and a market more responsive to evolving consumer needs.

Unfortunately, data-driven technologies have also powered a new generation of consumer scams. Digital technologies help scammers sift through potential targets so that their unscrupulous offerings appear on the screens precisely of those individuals most likely to fall victim. Often that means leveraging data to develop customized scams or to identify and target vulnerable populations.

For example, once a website learns that a consumer is shopping for a particular item, a common tactic is to advertise a fake, limited-time sale for the specific item that the customer is currently shopping for. In truth the item is always available at the specified price, but by recharacterizing the offer as a limited-time sale, the site creates a false sense of urgency. To increase pressure and augment the technique's power, a site might also include a fake countdown timer or fake inventory counter. Other techniques target consumers using fake news stories, endorsements from their favorite celebrities or politicians, or fake posts from friends and social-media contacts. By tying their products to a consumer's social network, known news sources, or media personalities, scammers can overcome a consumer's ordinary caution toward unknown products and increase the likelihood that a scam will succeed.

## III. LEGISLATIVE AND REGULATORY RESPONSES

The wave of AI-powered online scams has lawmakers around the world scrambling for solutions. What makes the problem especially challenging is that, in an important sense, there is nothing new about online consumer deception at all. Fraudsters and targeted scams are as old as time. What is new and dangerous about online scams is not any special power to deceive, but their unprecedented efficiency. Digital marketing technologies bring the marginal cost of scamming one more consumer near to zero, just the cost of one more targeted ad. (Dickinson [2024]).

As in other contexts, seller reputational constraints and the principle of caveat emptor serve as the most important defenses for consumers. To aid consumers in protecting their interests, the law also sometimes mandates that sellers disclose certain categories of information, often in a specified format. For example, FTC advertising guidelines require social-media influencers to disclose any "material connection" they have with a brand (such as the brand paying them or providing free or discounted products) so that consumers can weigh that information when considering an influencer's endorsement.

One recurring question in consumer protection law is to what extent lawmakers should take matters a step further, by not only mandating disclosures that facilitate voluntary bargaining, but by also banning certain categories of transactions even if voluntarily agreed to. For example, long-standing usury laws in many jurisdictions prohibit lenders from offering and borrowers from accepting loans at an interest rate greater than some specified limit. Proponents of such laws reason that interest above a certain threshold is simply unfair. Critics, by contrast, reason that each individual is best positioned to evaluate her own circumstances and worry that interest-rate limits undermine the ability of those with poor credit histories to obtain financing for business opportunities.

A modern version of the debate is playing out in the online world. With the digital age have come new business models, often powered by consumer data. Consider two of the industry's leaders, Google and Facebook, which offer their digital products to millions of users worldwide, all for free. App makers offer high-quality digital products for free in exchange for permission to collect their users' data. Instead of charging their users fees directly, the companies analyze their users' web-browsing and smartphone usage data for insights regarding the products consumers are most likely to be interested in. They can then fund their operations by selling targeted advertising services to other companies interested in promoting their products to specific types of users.

Because such business models require consumer data to operate, a consumer installing a new app or creating a new online account will almost invariably be presented with contractual terms requesting her consent to the company's collection and use of her data. Data collection powers the apps so many have come to love and rely on; but it can also feel invasive to receive highly targeted ads. What is more, the same targeted advertising techniques used to present users with relevant ads can also be misused by scammers, for example to target vulnerable communities such as nonnative language speakers or to tailor scams to particular communities, for example, a fake celebrity endorsement by Kim Kardashian presented to users determined to be among her fans. (Dickinson [2024]).

Concern over the use and misuse of consumer data has spurred lawmakers to adopt major legislative reforms over the last decade. Most significant is the EU's General Data Protection Regulation (GDPR), adopted in 2016, which, beyond requiring entities to disclose to consumers how their data will be used and to obtain consent to such usage, also guarantees consumers various rights regarding the data collected about them. (Yun [2024]). Among other rights, GDPR guarantees consumers the "right to access" and "right to data portability," meaning that consumer can demand a copy of whatever data an entity has related her, and the "right to erasure," also known as the "right to be forgotten," which, with various exceptions, entitles the consumer to demand that an entity delete whatever information it has concerning her. Following the State of California's adoption of the California Consumer Privacy Act in 2018, more than a dozen U.S. states have also enacted comprehensive consumer data privacy laws that include many of the same provisions.

Another related issue occupying lawmakers' attention is what, if anything, should be done to address so-called "dark patterns"—app and website user interface designs that persuade, trick, or coerce users into making purchases, sharing browsing data, or making whatever other choices their designers prefer. Common complaints include subscription services that require users to navigate multiple screens to cancel and sales offers that can only be declined by clicking buttons with such labels as "No Thanks, I Don't Like Saving Money." (Dickinson [2023]).

The most significant legislative action to date regarding dark patterns is the EU's Digital Services Act (DSA), Articles 25 of which prohibits any user interface design that deceives, manipulates, or "otherwise distorts or impairs" users' ability to make free and informed decisions. Insofar as the DSA bars sellers from misrepresenting their products and services, it merely reiterates long-standing prohibitions on fraud and deceptive trade practices. Yet by also prohibiting the "manipulation" of consumers or the "impairment" of their decision making, the DSA sweeps more broadly, to bar not only offers likely to mislead consumers but also truthful but especially persuasive offers that, in the view of regulators, consumers would be better off declining.

## IV. CONCLUSION

The DSA, GDPR, CCPA, and other recent laws governing online transactions show a trend toward increased governmental scrutiny of digital transactions. Although the historical goal of consumer protection law has been to ensure that transactions are voluntary and untainted by deception, lawmakers in the digital age have become increasingly willing to prohibit even some voluntary, nondeceptive transactions. Although phrased as consumer data "rights," GDPR and CCPA function, in effect, to prohibit a consumer from irrevocably agreeing to share her data. She cannot so agree even if she wants to, for the data privacy laws perpetually reserve for her a right later to demand erasure of the data. Similarly, the DSA's vague prohibitions on consumer "manipulation" or "impairment" of choice cast an uncertain shadow over transactions historically considered voluntary, if it seems a consumer might have or should have chosen differently in a less persuasive or differently structured choice environment.

As with most things, there is no single solution to online consumer protection, just a set of options and tradeoffs. The best available data on the effects of GDPR in the EU, for example, show a reduction in consumer data collection and web tracking, but corresponding increases in search costs for consumers, who face greater difficulty finding products; a reduction in new smartphone app creation, given that restrictions on the use of consumer data might prevent companies from covering their costs; and increased market concentration, as without the benefit of targeted ads small companies find it more difficult to compete with established brands. (Yun [2024]).

Every piece of consumer protection legislation past and present raises similar tradeoffs: Usury laws prevent some consumers from entering unwise arrangements, while also preventing some low-cost borrowers from obtaining financing. (CFPB

Report at ch. 5). CCPA, like GDPR, limits the scope of permissible data collection arrangements, increasing consumer data privacy, while reducing the relevance of the ads they see and impeding innovation by undermining a key source of revenue. (Yun [2024] at 122). And prohibitions on "dark patterns" streamline some app and website functionality while increasing legal uncertainty and reducing companies' willingness to experiment with innovative app designs. (Dickinson [2023] at 1635). The key concern for digital consumer protection is thus to inventory the regulatory avenues available and to consider the tradeoffs of their implementation.